\documentclass[5p,twocolumn,times,number]{elsarticle}

\usepackage{graphicx}
\usepackage{amsmath}   

\begin{document}

\begin{frontmatter}

\title{The Mu2e calorimeter: quality assurance of production crystals and SiPMs}

\author[a]{N.~Atanov}
\author[a]{V.~Baranov}
\author[a]{J.~Budagov}
\author[e,f]{D.~Caiulo}
\author[e]{F.~Cervelli}
\author[b]{F.~Colao}
\author[b]{M.~Cordelli}
\author[b]{G.~Corradi}
\author[a]{Yu.I.~Davydov}
\author[e,1]{S.~Di Falco\corref{cor}}
\ead{stefano.difalco@pi.infn.it}
\author[b,i]{E.~Diociaiuti}
\author[e,f]{S.~Donati}
\author[b,j]{R.~Donghia}
\author[c]{B.~Echenard}
\author[b]{S.~Giovannella}
\author[a]{V.~Glagolev}
\author[h]{F.~Grancagnolo}
\author[b]{F.~Happacher}
\author[c]{D.G.~Hitlin}
\author[b,d]{M.~Martini}
\author[b]{S.~Miscetti}
\author[c]{T.~Miyashita}
\author[e]{L.~Morescalchi}
\author[g]{P.~Murat}
\author[e]{E.~Pedreschi}
\author[k]{G.~Pezzullo}
\author[c]{F.~Porter}
\author[e]{F.~Raffaelli}
\author[b,d]{M.~Ricci}
\author[b]{A.~Saputi}
\author[b]{I.~Sarra}
\author[e]{F.~Spinella}
\author[h]{G.~Tassielli}
\author[a]{V.~Tereshchenko}
\author[a]{Z.~Usubov}
\author[a]{I.I.~Vasilyev}
\author[c]{R.Y.~Zhu}

\cortext[cor]{Corresponding author}

\address[a]{Joint Institute for Nuclear Research, Dubna, Russia}
\address[b]{Laboratori Nazionali di Frascati dell'INFN, Frascati, Italy}
\address[c]{California Institute of Technology, Pasadena, United States}
\address[d]{Universit\`a ``Guglielmo Marconi'', Roma, Italy}
\address[e]{INFN Sezione di Pisa, Pisa, Italy}
\address[f]{Dipartimento di Fisica dell'Universit\`a di Pisa, Pisa,
 Italy}
\address[g]{Fermi National Laboratory, Batavia, Illinois, USA}
\address[h]{INFN Sezione di Lecce, Lecce, Italy}
\address[i]{Dipartimento di Fisica dell'Universit\`a di Roma Tor Vergata, Rome, Italy}
\address[j]{Dipartimento di Fisica dell'Universit\`a degli Studi Roma Tre, Rome, Italy}
\address[k]{Yale university, New Haven, USA}

\begin{abstract}

The Mu2e calorimeter is composed of two disks each containing 1348 pure CsI
crystals, each crystal read out by two arrays of 6x6 mm$^2$ monolithic SiPMs.
The experimental requirements have been translated in a series of technical
specifications for both crystals and SiPMs. 
Quality assurance tests, on first crystal and then SiPM production batches, confirm the performances of preproduction samples previously assembled in a calorimeter prototype and tested with an electron beam. The production yield is sufficient to allow the construction of a calorimeter of the required quality in the expected times.

\end{abstract}

\begin{keyword}
Calorimeter \sep Crystals \sep SiPM

\PACS 29.40.Cs \sep 29.40.Gx    
\end{keyword}

\end{frontmatter}

\section{Introduction}

The Mu2e experiment\cite{TDR} at Fermilab will search for the conversion of a muon to an electron in the field of an Aluminum atom improving by 4 orders of magnitude all previous sensitivities on this kind of Charged Lepton Flavor Violating process. The experiment consists of three superconducting solenoids containing the production and the stopping target, the tracker and the electromagnetic calorimeter and an external system used to veto cosmic rays. 
The calorimeter \cite{ECAL} helps the tracker in the identification of converted electrons (CE), building an efficient trigger and improving the track reconstruction efficiency. According to Monte Carlo simulations, the required calorimeter performances on CE are as follows: an energy resolution better than 10\%, a time resolution better than 500 ps and a position resolution better than 1 cm.  
The calorimeter consists of two disks with 674 pure CsI crystals each; each crystal is coupled in air to two arrays of 2$\times$3 UV-extended Hamamatsu SiPMs. Quality criteria for crystals and SiPM arrays have been determined by a test campaign on pre-production samples \cite{crystalQC,SiPMprepro}. A prototype built with crystals and SiPMs fulfilling the quality criteria has been exposed to a beam  to verify, with the help of Monte Carlo simulation, that the energy and time resolutions meet the requirements of the experiment\cite{donghia}. 
The results of the first quality assurance tests are presented here.

\section{Crystals quality assurance}

The quality criteria for CsI crystals are the following: dimensions within the 100 $\mu$m mechanical tolerance(MT); light output (LO), measured with a 2'' UV extended EMI PMT, higher than 100 p.e./MeV; longitudinal response uniformity (LRU) better than 5\%;  slow component due to crystal impurities less than 25\%; LO reduction after a ionizing dose of 1 kGy lower than $40\%$; negligible reduction of LRU and LO after a fluence of $9\times 10^{11}$ neutrons/cm$^2$ 1 MeV equivalent; radiation induced noise (RIN) at 18 mGy/h equivalent to less than 600 KeV.

\begin{figure}
\centering
\includegraphics[width=0.99\linewidth]{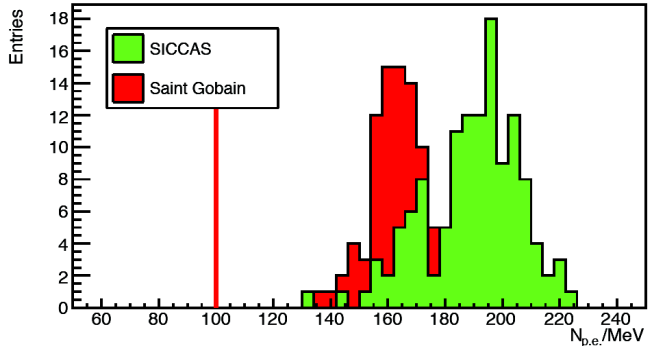}
\caption{Light output(LO) of first Saint-Gobain and SICCAS production crystals: all have LO$>$100 p.e./MeV. Histograms are stacked.  }
\label{fig:lightyield}
\end{figure}

An automated station measures LO (fig.\ref{fig:lightyield}) and LRU moving  a $^{22}$Na source along the crystal. Another station measures RIN by radiating 6 crystals/batch with a $^{137}$Cs source delivering 0.42 mGy/h. Radiation hardness will be checked at Caltech on a random subsample of each production batch. 
Two vendors have been chosen for crystal production. 
Of the 322 SICCAS crystals tested so far only 5\% have been rejected because of MT and 2\% because of LRU. About 44\% of the 100 Saint-Gobain crystals have been rejected because of MT. An on going revision of vendor procedures is expected to improve this yield. 

\section{SiPM quality assurance}

Each SiPM array consists of two series of 3 monolithic 6$\times$6 mm$^2$ cells connected in parallel. 
The quality criteria for SiPM arrays are the following:
for the 6 cells in the array the breakdown voltage ($V_b$) spread must be better than 0.5\% and the  dark current ($I_{dark}$) spread must be better than 15\%; the gain at the operating voltage ($V_{op}=V_b+3V$) must be higher than $10^6$ when measured with a 150 ns gate; photon detection efficiency (PDE) must be higher than 20\%; after an irradiation with $3\times 10^{11}$ neutrons/cm$^2$ 1 MeV equivalent, $I_{dark}$ at 0$^\circ$ C  must be lower than 2 mA; the mean time to failure (MTTF) must be higher than $6\cdot10^5$ h.

\begin{figure}
\centering
\includegraphics[height=0.53 \linewidth]{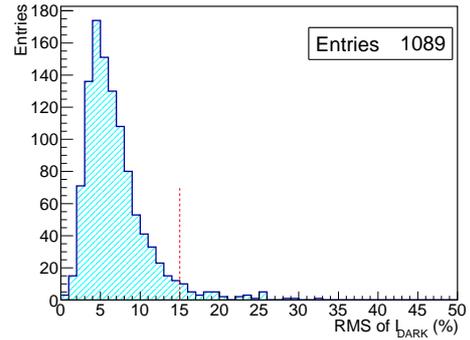}
\caption{Dark current spread for the six cell of SiPM arrays measured at 20$^o$C: about 3\% of the arrays having an RMS$>$15\% have been rejected so far.}
\label{fig:Idark}
\end{figure}

Hamamatsu delivers a batch of $\sim$300 of SiPM arrays per month. SiPM characteristics are tested at the SIDET facility at Fermilab following the procedures described in detail in \cite{SiPMtest}.     After a dimensional check, 25 SiPM arrays (20+5 used as reference) are tested at the same time at 3 different temperatures (-10$^\circ$C, 0$^\circ$C and 20$^\circ$C). The $I_{dark}$ vs $V_{bias}$ curve, acquired by an automatic LabView program, is used to determine $V_b$. The gain$\times$PDE at $V_{op}$ is obtained by comparing the response to a green led light with the one of the reference SiPMs. In the mean time, an accelerated aging test at 65$^\circ$C is performed on 15 arrays per batch to ensure the required MTTF. Radiation hardness to neutrons is checked on 5 arrays per batch at the Helmholtz Zentrum Dresden Rossendorf (Germany). 
Of the $\sim$1500 Hamamatsu production SiPM tested so far, less than 4\% have been rejected because of dark current spread (RMS in fig. \ref{fig:Idark}).

\section{Summary and conclusions}
Production yield for Hamamatsu SiPMs and SICCAs crystals is excellent. Saint-Gobain crystals yield is limited to $\sim$50\% because of MT, but improvements are expected in the near future. No major problems are envisaged at the moment in the construction of an high quality Mu2e calorimeter.
 
\section*{Acknowledgments}

We are grateful for the vital contributions of the Fermilab staff and the technical staff of the participating institutions.
This work was supported by the US Department of Energy; 
the Italian Istituto Nazionale di Fisica Nucleare;
the Science and Technology Facilities Council, UK;
the Ministry of Education and Science of the Russian Federation;
the US National Science Foundation; 
the Thousand Talents Plan of China;
the Helmholtz Association of Germany;
and the EU Horizon 2020 Research and Innovation Program under the Marie Sklodowska-Curie Grant Agreement No.690385. 
Fermilab is operated by Fermi Research Alliance, LLC under Contract No.\ De-AC02-07CH11359 with the US Department of Energy, Office of Science, Office of High Energy Physics.
%


\end{document}